\documentclass[usenatbib]{mn2e}
\usepackage{graphicx,times}
\usepackage{bm}
\usepackage{epsf}
\usepackage{amsmath}
\usepackage{amssymb}

\def\br{{\bf r}}

\def\n{\noindent}

\def\gg{{}_s\Gamma}
\def\ggp{{}_{s'}\Gamma'}
\def\ggpp{{}_{s''}\Gamma''}
\def\ggppp{{}_{s'''}\Gamma'''}

\def\sg{\Gamma}
\def\sgp{\Gamma'}
\def\sgpp{\Gamma''}
\def\sgppp{\Gamma'''}

\def\ad{{_\delta}}

\def\bk{{\bf k}}

\def\2p{{(2\pi)^2}}

\def\be{\begin{equation}}
\def\ee{\end{equation}}

\def\beq{\begin{equation}}
\def\eeq{\end{equation}}

\def\ben{\begin{eqnarray}}
\def\een{\end{eqnarray}}

\def\oh{{\hat\Omega}}
\def\nn{{\nonumber}}

\newcommand{\beqa}{\begin{eqnarray}}
\newcommand{\eeqa}{\end{eqnarray}}

\newcommand{\edth}{\,\eth\,}

\newcommand{\calZ}{{\cal Z}}

\date{\today,~ $ $Revision: 0.9 $ $}

\begin{document}

\onecolumn

\title[Statistics of Weak Lensing Shear and Flexion]
{Higher-order Statistics of Weak Lensing Shear and Flexion}

\author[Munshi et al.]
{Dipak Munshi$^{1,2}$, Joseph Smidt$^{3}$, Alan Heavens$^{1}$, Peter Coles$^2$, Asantha Cooray$^{3}$\\
$^{1}$Scottish Universities Physics Alliance (SUPA), Institute for Astronomy, University of Edinburgh, Blackford Hill,  Edinburgh EH9 3HJ, UK \\
$^{2}$School of Physics and Astronomy, Cardiff University, Queen's
Buildings, 5 The Parade, Cardiff, CF24 3AA, UK \\
$^{3}$ Department of Physics and Astronomy, University of California, Irvine, CA 92697, USA \\}

\maketitle

\begin{abstract}
Owing to their more extensive sky coverage and tighter control on
systematic errors,  future deep weak lensing surveys should provide
a better statistical picture of the dark matter clustering beyond
the level of the power spectrum. In this context, the study of
non-Gaussianity induced by gravity can help tighten constraints on
the background cosmology by breaking parameter degeneracies, as well
as throwing light on the nature of dark matter, dark energy or
alternative gravity theories. Analysis of the shear or flexion
properties of such maps is more complicated than the simpler case of
the convergence due to the spinorial nature of the fields involved.
Here we develop analytical tools for the study of higher-order
statistics such as the bispectrum (or trispectrum) directly using
such maps at different source redshift. The statistics we introduce
can be constructed from cumulants of the shear or flexions,
involving the cross-correlation of  squared and cubic maps at
different redshifts. Typically, the  low signal-to-noise ratio
prevents recovery of the bispectrum or trispectrum mode by mode. We define
power spectra associated with each multispectra which compresses
some of the available information of higher order multispectra. We
show how these can be recovered from a noisy observational data even
in the presence of arbitrary mask, which introduces mixing between
{\em Electric} (${\rm E}$-type) and {\em Magnetic} (${\rm B}$-type)
polarization, in an unbiased way. We also introduce higher order
cross-correlators which can cross-correlate lensing shear with
different tracers of large scale structures.
\end{abstract}

\section{Introduction}

Weak lensing surveys play an important role as cosmological probes
complementary to Cosmic Microwave Background (CMB) surveys and
large-scale galaxy surveys. In principle, they can probe evolution
of the dark matter power spectrum at a moderate redshifts in an
unbiased way; for a recent review, see \cite{MuPhysRep08}. However,
the first weak lensing measurements were published within the last
decade \citep{BRE00,Wittman00,KWL00,Waerbeke00} so this is a young
field. There has been a tremendous progress in the last decade since
the first measurements, on various fronts, including analytical
modeling, technical specification and the control of systematics.
Weak lensing probes cosmological perturbations on smaller angular
scales where the perturbations are both nonlinear and non-Gaussian.
The two-point correlation function (or, equivalently, the power
spectrum) of weak lensing is weakly sensitive to the cosmological
constant $\Omega_\Lambda$. It also depends only on a degenerate
combination of amplitude of matter power spectrum $\sigma_8$ and the
matter density parameter $\Omega_M$. This degeneracy can be broken
by use of the three-point correlation function which is diagnostic
of the non-Gaussianity induced by gravity
\citep{Vil96,JainSeljak97}.

The higher order statistics needed to analyze weak lensing in detail
are however prone to the effects of noise due to the intrinsic
ellipticity distribution, finite number of galaxies (shot noise) and
sample (or cosmic) variance owing to the  partial sky coverage.
Ongoing and planned weak lensing surveys, such as the CFHT legacy
survey{\footnote{http://www.cfht.hawai.edu/Sciences/CFHLS/}},
Pan-STARRS  {\footnote{http://pan-starrs.ifa.hawai.edu/}}, the Dark
Energy Survey,  and further in the future, the Large Synoptic Survey
Telescope {\footnote{http://www.lsst.org/llst\_home.shtml}}, JDEM
and Euclid will provide a wealth of information in terms of mapping
the distribution of mass and energy in the universe. A large
fractional sky coverage as well as accurate photometric redshift
determination will help to probe higher order correlation functions
at a  higher signal-to-noise(S/N) then is possible at present (see
e.g. \cite{Pen03}).

Almost all previous studies concerning weak lensing have used a real
space description and employed correlation function of various
orders to probe underlying mass clustering. Future surveys will
cover a good fraction of the sky, if not the entire sky. This
motivates us to use a harmonic description and employ the
mutispectra - which are simply the Fourier representations of the
higher order correlation functions. Weak lensing observations
require masks with complicated topology to deal with presence of
foreground objects. Our formalism in harmonic space can deal with
arbitrary mask. The formalism described here is completely general
and can deal with fields with arbitrary spins including shear and
flexions.

In the absence of information concerning the redshifts of the source
galaxies, traditional weak lensing studies have tended to operate in
projection or in 2D slices. Indeed, before the advent of 3D weak
lensing, almost all studies were carried out in such a way
\citep{JSW00}. Due to lack of all-sky coverage most studies also
used a ``flat sky'' approach (\citet{MuJai01,Mu00,MuJa00}) which has
led to study of lower order non-Gaussianity using convergence
$\kappa$ as well as shear $\gamma$ \citet{MuJai01,Valageas00,
MuVa05,VaMuBa04,VaMuBa05}. As the next stage of development
tomographic approaches went beyond projected survey by binning
galaxies in redshift slices which can further tighten the
constraints \citet{TakadaWhite03,TakadaJain04}. More recently, the
use of photometric redshifts to study weak lensing in three
dimensions was introduced by \citet{Heav03}. It was later developed
by many authors \citet{HRH00, HKT06, HKV07, Castro05}, and was shown
to be a vital tool in constraining dark energy equation of state
\citep{HKT06}, neutrino mass \citep{Kit08} and many other
possibilities. In our present study we propose estimators for
non-Gaussianity using projected data.

In most cosmological studies, the power spectrum remains the most
commonly used statistical probe and its evolution and
characterization remain the best studied. As has already been
pointed out, however, higher order statistics can help to break
degeneracies e.g. in $\Omega_M$ and $\sigma_8$, but are more
difficult to probe observationally (see e.g.
\citet{BerVanMell97,JainSeljak97, Hui99,Schneider98,TakadaJain03})
given the low signal to noise associated with them. As a result most
commonly used higher order probes compress the available information
to a simple number (e.g. ``skewness'' or ``kurtosis'') which are
one-point statistics \citep{Pen03}. The higher order correlation
functions have been detected observationally
\citep{BerVanMell97,BerVanMell02} and future larger samples are
expected to improve the statistical situation still further. A
recent study by \citet{Mu3D10} has underlined the usefulness of
working with two-point correlations (or their Fourier analog the
power spectrum) at each order. At the level of the bispectrum, a
single power spectrum tends to compress some of the available
information; there is more than one degenerate power spectrum
associated with the higher order multispectra. We extend the recent
study by \citet{Mu3D10} and provide systematic results at each order
for spin$-2$ fields, thus generalizing our previous results for
spin-$0$ fields such as the convergence. Understanding higher order
statistics also provide an indication of the scatter associated with
estimation of lower order statistics \citet{TakadaJain09}. Their
results extend the formalism  developed in \citet{Heav03} and
\citet{Castro05} and we follow the all-sky expansion introduced in
weak lensing studies by \citet{Stebbins96}.

Statistics of the shear $\gamma$ or convergence $\kappa$ seen in a
lensing map are sensitive to the statistics of underlying density
contrast $\delta = {\delta \rho_b / \rho_b}$ of background density
$\rho_b$. An accurate understanding of the gravitational clustering
is therefore essential for modeling the weak lensing statistics. At
present we lack a detailed analytical model of the non-linear
gravitational clustering. In the absence of a clear analytical
picture, beyond the perturbative regime the modeling of the higher
order statistics are done using a hierarchical ansatz (see e.g.
\citet{Fry84,Schaeffer84, BerSch92,SzaSza93, SzaSza97, MuBaMeSch99,
MuCoMe99a, MuCoMe99b, MuMeCo99, MuCo00, MuCo02, MuCo03}). The
hierarchical ansatz assumes a factorisable model for the higher
order correlation functions. Different hierarchical ansatze differs
the way they assign amplitudes to the various tree amplitudes. We
will employ a very generic form of the hierarchical ansatz which has
already been tested in modeling of projected statistics. The method
is flexible enough to allow for a more specific prescription. Other
approaches to model nonlinear gravity include the halo models
\citep{CooSeth02} which has also been employed in weak lensing
studies.

In a recent paper, \citet{Mu3D10} advocated the use of higher order
cumulants and their correlators of convergence $\kappa$ along with
their associated power spectra for studying dark matter clustering
beyond power spectra in 3D. The aim of the present paper is to
extend that study to fields with non-zero spin (e.g. shear). Shear
can directly be used from observed data to study the dark matter
clustering beyond the usual approximation of Gaussianity. Although
in this study we primarily focus on shear, the analytical results
are directly applicable for higher order spin fields sometime used
in weak lensing known as {\it flexions}. Previous studies have also
focused on the  cumulant correlators of the shear $\gamma$ and
$\kappa$ \citep{BerVanMell97,BerVanMell02,BerMellWaer03}. While
cumulant correlators for convergence fields can be probed in a
relatively straight forward manner, even to arbitrary order
\citep{Mu00}, the analogous studies are more involved for shear
fields. There have been some studies involving smoothed $M_{ap}$
statistics to study cumulant correlators from shear map by using
compensated filters \citep{MuVaCross05}, but in this present paper
we develop a more systematic approach to probe higher order
statistics directly through spinorial objects. The focus of the
present study is to develop tools in the Fourier domain and to use
the power spectra associated with the multispectra as opposed to
cumulant correlators. Given the non-trivial topology of weak lensing
surveys correlation functions indeed are a good option to work with,
but, as we will show, the power spectra we develop can also be
estimated in the presence of arbitrary mask and noisy data.

The paper is arranged as follows. In \textsection2 we discuss the
basic formalism and introduce some terminology and notation. In
\textsection3, we introduce the power spectra related to bispectra
and trispectra for various combinations of the convergence, shear as
well as flexion fields. In \textsection4 we focus on statistical
description of underlying matter distribution and relate them to
power spectra defined in previous sections for weak lensing
observables. Finally \textsection5 is reserved for discussions and
future prospects.

\begin{figure}
\begin{center}
{\epsfxsize=6. cm \epsfysize=6. cm {\epsfbox[301 425 590 713]{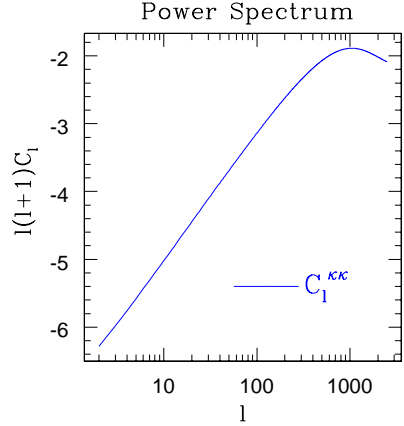}}}
\end{center}
\caption{The plots correspond to the power spectrum of convergence $C_l^{\kappa\kappa}$. 
The source redshift is $z_1=z_2=1$. See text for more details.}
\label{fig:ps}
\end{figure}

\section{Convergence, Shear and Flexions}

Among the observables including in weak lensing studies are the
convergence $\kappa(\oh)$, shear $\gamma(\oh)$ and flexions ${\cal
F}(\oh), {\cal G}(\oh)$. The harmonic decomposition needed in
all-sky calculations involves spin-weight spherical harmonics for
such  objects. We introduce the basic notations in this section
following \cite{Castro05}. We will ignore the radial dependence for
this work as we mainly concentrate on projected surveys. The results
presented here will extend the results in \citep{Mu3D10} wherein
only scalar fields were considered. Though the analysis here follows
full spherical decomposition it is indeed possible to choose a
rectangular coordinate system which can simplify the calculation for
surveys observing a small patch of the sky.

\subsection{Spin-weight Spherical Harmonics}

\n We start by introducing the spin-weight spherical harmonics
${}_sY_{lm}(\oh)$ that can be expressed in terms of the Wigner-D
function (see e.g. \cite{VMK88,PenRind84} for a detailed
discussion). They form a complete and orthonormal set of function on
the surface of the celestial sphere. Any spin-weighted function
${}_sf(\oh)$ can be decomposed into harmonics ${}_sf_{lm}$ using
spin-weight spherical harmonics:

\be
{}_sY_{lm}(\oh) = \sqrt \frac{2l+1}{4 \pi} D^l_{-s,m}(\theta,\phi,0); \qquad {}_sf(\oh)= \sum_{lm} {}_sY_{lm}(\oh) {}_sf_{lm}. \\
\ee

\n The spin-weighted spherical harmonics ${}_sY_{lm}(\oh)$ (defined
only for $|s|>l$ )can be seen as a generalization of the scalar,
vector and tensor, spherical harmonics \citep{VMK88} satisfying the
following orthogonality and completeness relations:

\be
\sum_{lm} {}_sY_{lm}(\oh){}_{s'}Y_{lm}(\oh') = \delta_D(\oh-\oh'); \qquad \int d\oh~{}_sY_{lm}(\oh) {}_{s'}Y_{l'm'}(\oh) = \delta^K_{ss'}\delta^K_{ll'}\delta^K_{mm'}.
\label{spherical}
\ee

\n Here $\delta^K$ is the Kronecker delta function and $\delta_D$ is
a Dirac delta function. The spin raising operator ``{\em edth}'',
$\edth$ and  its complex conjugate, the spin lowering operator
$\bar\edth$ are defined through their action on spin spherical
harmonics which can be used to define spin-weighted harmonics
${}_sY_{lm}(\oh)$ from ordinary spherical harmonics $Y_{lm}(\oh)$.
The spin spherical harmonics ${}_sY_{lm}(\oh)$ are the eigen
functions of the operator $\edth\bar\edth$. Note that $\edth$ can be
thought also as an effective covariant derivative on the surface of
the sphere which generates quantities related to the various spins:

\be \edth {}_sY_{lm} = [(l-s)(l+s+1)]^{1/2} {}_{s+1}Y_{lm}; \qquad
\bar\edth {}_sY_{lm} = [(l+s)(l-s+1)]^{1/2} {}_{s-1}Y_{lm}. \ee \n
This also implies that $\edth\bar\edth {}_sY_{lm} = [(l-s)(l+s+1)]
{}_{s}Y_{lm}$ and $\edth\bar\edth {}_sY_{lm} = [(l+s)(l-s+1)]
{}_{s}Y_{lm}$. The conversion between convergence and shear
harmonics used the following expression which we will also be using
later:

\be
\edth\edth Y_{lm}(\oh) = \sqrt {(l+2)! \over (l-2)! } {}_2Y_{lm}(\oh); \qquad\qquad
\bar\edth\bar\edth Y_{lm}(\oh) = \sqrt  {(l+2)! \over (l-2)! } {}_{-2}Y_{lm}(\oh).
\label{updown}
\ee

\n The operators $\edth$ and $\bar\edth$ are convenient for
expressing the spin weight objects which are coordinate frame
dependent in terms of ones which are frame independent or scalars.
All invariant differential operators on the sphere may be expressed
in terms of $\edth$ and $\bar\edth$.

The following overlap integrals will be useful in our future derivations which will be used to
derive the expressions for bispectrum and trispectrum related power spectrum.

\beq
\int {}_{s}Y_{lm}(\oh){}_{s'}Y_{l'm'}(\oh){}_{s''}Y_{l''m''}(\oh) d\oh =  S_{ll'l''}
\left ( \begin{array}{ c c c }
     l & l' & l'' \\
     m & m' & m''
  \end{array} \right)
\left ( \begin{array}{ c c c }
     l & l' & l'' \\
     -s & -s' & -s''
  \end{array} \right); \qquad S_{ll'l''} \equiv \sqrt{(2l+1)(2l'+1)(2l''+1)\over 4\pi}.
\label{eq:harmonics1}
\eeq

\n The above result can be cast in following form which is useful
for expressing integrals of more than three spin spherical harmonics
in terms of Wigner $3j$ symbols when dealing with trispectrum or
 multispectra of even higher order.

\beq
{}_{s}Y_{lm}(\oh){}_{s'}Y_{lm}(\oh) = \sum_{LSM}{}_SY^*_{LM} S_{Ll'l''}
\left ( \begin{array}{ c c c }
     L & l' & l'' \\
     M & m' & m''
  \end{array} \right)
\left ( \begin{array}{ c c c }
     L & l' & l'' \\
     -S & -s' & -s''
  \end{array} \right).
\label{eq:harmonics2}
\eeq

\n
These results are generalizations of similar results often used in the context of scalar (spin-$0$)
harmonics which can be recovered by setting $s=s'=s''=0$. In a small patch of sky, a rectangular
coordinate system can be introduced which can simplify some
of the calculations and might be useful for surveys with small sky coverage. A detailed
analysis will be presented elsewhere (Munshi et al. (2010) in preparation).

\subsection{Harmonic decomposition of Convergence and Shear}

We start by defining the complex shear $\gamma$ in terms of the
individual shear components $\gamma_1$ and $\gamma_2$ at a given
angular position $\Omega=(\theta,\phi)$ and at a radial
line-of-sight distance $|\br|=r$ (see \cite{MuPhysRep08} for a more
detailed discussion and definitions). We will closely follow the
notations of \citep{Castro05} where possible. We will not Fourier
decompose in the radial direction. The harmonic decomposition is
performed only the surface of the celestial sphere

\be
\gamma_{\pm}(\oh,r) = \gamma_1(\oh,r) \pm i \gamma_2(\oh,r).
\ee

\n It is customary to define a complex lensing potential $\phi(r) =
\phi_E(r)+i\phi_B(r)$ which is related to the line-of-sight
integration of the peculiar gravitational potential $\phi$ and to
define the shear in terms of spin-derivatives $\eth$ and its
conjugate $\bar\eth$ of $\phi$. The action of $\eth$ is to raise the
spin whereas $\bar\eth$ reduces the {\it spin} of the object it is
acting on. This operator was first introduced by
\cite{VMK88,PenRind84} on the surface of the sphere in order to
define the now widely used {\it spin-weight $s$ spherical harmonics}
which live on the surface of a sphere

\be
\gamma({\bf r}) = {1 \over 2}\eth\eth [ \phi_E(\br)+i\phi_B(\br)]; \qquad
\gamma^*({\bf r}) = {1 \over 2}{\bar\eth}{\bar\eth}[ \phi_E(\br)-i\phi_B(\br)].
\qquad
\ee

\n Explicit expressions for these spin-lowering and raising
differential operators are given in \cite{Castro05}. While acting on
a complex scalar potential $\phi$, the fields $\gamma({\bf r})$
generates the $\gamma({\bf r})$ and its conjugate which are of spin
$+2$ and $-2$ respectively. In later sections we will use the
generalized symbol ${}_s\Gamma$ for general spin fields which will
include products of shear fields as well as higher derivative spin
fields such as flexions. In our current notation ${}_2\Gamma =
\gamma$ and ${}_{-2}\Gamma = \gamma^*$:

\be
{}_{2}\Gamma(\br) \equiv \gamma(\br) = {1 \over 2}\eth\eth \phi(\br); \qquad {}_{-2}\Gamma(\br) \equiv \gamma^*(\br) = {1 \over 2}\bar\eth\bar\eth \phi^*(\br).
\ee

\n
The individual shear components $\gamma_1(\br)$ and $\gamma_2(\br)$ can be expressed in terms of a
complex lensing potential $\phi(\br) =\phi_E(\br) + i\phi_B(\br)$. The magnetic part of the potential $\phi_B(\br)$ will take contribution mainly from systematics and electric part corresponds to pure lensing contribution

\be
\gamma_1(\br)= {1 \over 4}(\edth\edth+\bar\edth\bar\edth)\phi(\br); \qquad
\gamma_2(\br) = -{i\over 4}(\edth\edth - \bar\edth\bar\edth)\phi(\br); \qquad \kappa(\br) =
{1 \over 4}(\edth\bar\edth+\edth\bar\edth)\phi(\br).
\ee

\n
The harmonics of $\gamma$ and $\gamma^*$ can be expressed in terms of the harmonics coefficients of $\Phi_E$ and $\Phi_B$
denoted by $E_{lm}$ and $B_{lm}$ respectively.

\be
{}_2 \Gamma_{lm} = -[E_{lm}+ i B_{lm}];  \qquad  {}_{-2} \Gamma_{lm} = -[E_{lm} - i B_{lm}].
\ee

\n
The spinorial fileds can likewise be expanded in an appropriate basis which uses the spin-spherical harmonics
${}_sY_{lm}(\oh)$. For the case of ${}_2 \Gamma_{lm}$ we have the following expression:

\be
{}_{\pm2} \Gamma(\br) =  \sum_{l=0}^{\infty} \sum_{m=-l}^{l} \sqrt{(l+2)! \over (l-2)! } ~\phi_{lm} ~{}_{\pm2}Y_{lm}(r,\oh).
\ee

\n
We will be only dealing with 2D part of the spherical expansion as we will be dealing with projected surveys in this
work. The harmonics $E_{lm}$ and $B_{lm}$  are harmonic components of {\it Electric} (E) and {\it Magnetic} (B) fields respectively.
In the absence of B-mode the harmonics of 
the shear components are directly related to the harmonic component
of the Electric field $E_{lm}$. The harmonic transforms of the shear components and convergence are linked to
the lensing potential $\phi$ as follows:

\be
E_{lm}(k) = -{1 \over 2} \sqrt{(l+2)!\over (l-2)!} \phi_{lm}(k) \qquad \kappa_{lm}(k) = - {l(l+1)\over 2} \phi_{lm}(k).
\label{ref:shearkappa}
\ee

\n
These expression are useful in relating the bi- or trispectrum of various quantities in terms of the convergence ones.

\subsection{Harmonic Decomposition of Flexions}

Higher spin objects known as {\it flexions} are sometimes also used
to study weak lensing; they are related to derivatives of the shear
(\citet{GN02,GB05,Beacon06, BG05,Schneider08}). There are two flexions
which are typically used; $\cal F$ also known as the first flexion (spin -$1$)
and $\cal G$ which is also known as the second flexion (spin -$3$). The
combinations of these two flexions can specify the weak ``arciness''
of the lensed image and hence can quantify the lensing distortion beyond that
encoded in shear. Their relationship with the shapelet formalism
have been discussed at length \citep{Ref03,BJ02,RefBeac03}. Both
flexions have been used extensively in the literature for individual
halo profiles and also for the study of substructures
\citep{Beacon06}. Our aim here, as it is in the case of shear, is to
focus mainly on higher order statistics of these objects for generic
underlying cosmological clustering. We will do so by expressing the
bispectrum of the shapelets in terms of that of the convergence
$\kappa$, as we did for shear $\gamma$. The first and the second
flexions $\cal F$ and $\cal G$ can be derived from the lensing
potential $\phi(\br)$ \citep{Castro05} by the use of the lowering and raising operators:

\be
{\cal F}(\br) = {1 \over 6} \left (\bar\edth\edth\edth+\edth\bar\edth\edth + \edth\edth\bar\edth \right)\phi(\br); \qquad
{\cal G}(\br) = {1 \over 2}\bar\edth\bar\edth\bar\edth \phi(\br).
\ee

\n
Flexions have been used primarily to measure the galaxy-galaxy lensing to probe the
galaxy halo density profiles. Their cosmological use will depend on an accurate 
understanding  of gravitational clustering at small angular scales.
In Fourier space we will denote the harmonics of $\cal F$ and $\cal G$ by
${\cal F}_{lm}$ and ${\cal G}_{lm}$ and we can use the Eq.(\ref{updown}) 
to express them in terms of the $\phi_{lm}$.

\be
{\cal F}_{lm} = {1 \over 6} l^{1/2}(l+1)^{1/2}\left ( 3l^2 + 3l-2 \right )\phi_{lm}; \qquad
{\cal G}_{lm} = {1 \over 2} \sqrt {(l+3)! \over (l-3)!} \phi_{lm}
\label{eq:flex}
\ee

\n 
Though we will primarily be focusing on higher order statistics of
shear the results are equally applicable for flexions (which
generalize the concept of shear). In addition to the shear
statistics we will also consider a scalar field in our analysis
$\Psi$. We will leave this, which can represent a suitable large
scale tracer, arbitrary. The resulting analysis depends on the
multispectra involving both the weak lensing shear field and that of
the tracer fields. The flexions put more weight on smaller scales,
where the hierarchical ansatz is known to be more accurate, so the
formalism we develop here will also be suitable for them. It is also
possible to use our general results to construct estimators for
mixed bispectra involving shear and flexions.

However we should also point out that the modeling of intrinsic flexions 
of source galaxies, which is the main contribution
to the noise, is difficult. This will depend heavily on modeling
of galaxy shapes beyond the simplest description 
of their intrinsic shear. This uncertainty is expected to increase with 
survey depth. A detailed modeling and optimization of survey strategy
will be presented elsewhere.

\section{Higher Order Statistics of Weak Lensing Shear and Flexion}

The Higher order statistics encodes the departure from Gaussianity. The non-Gaussianity probed by any
cosmological observations can either be a primary one due to initial conditions or a secondary one which in case of
weak lensing is mainly induced by subsequent gravitational instability responsible for structure formation.
At lower redshift it is expected that the secondary non-Gaussianity
will dominate the primary one. The estimation of non-Gaussianity using higher order correlation functions
are typically dominated by noise. The higher order correlation functions or
their Fourier transforms (multispectra) contains a wealth of information. Some of these information
however can be degenerate due to large number of possible configurations in real or Fourier domain.
Typically one-point statistics e.g. skewness or kurtosis that are used compress all possible
information to a single number thereby increasing the Signal-to-Noise but reducing the information content.
As a next step, collapsed two-point objects (or cumulant correlators) which are higher order objects
are also used \citep{Mu00}. They are easier to recover from a noisy data than the complete correlation functions. The Fourier
transforms of these two-point cumulant correlators are the power spectrum associated with a
specific choice of the multispectra. There can be more than one power spectra for any specific choice of multispectra.
In this section we derive power spectra associated with bispectra and trispectra for shear as well as flexions.

\begin{figure}
\begin{center}
{\epsfxsize=6. cm \epsfysize=6. cm {\epsfbox[25 431 305 712]{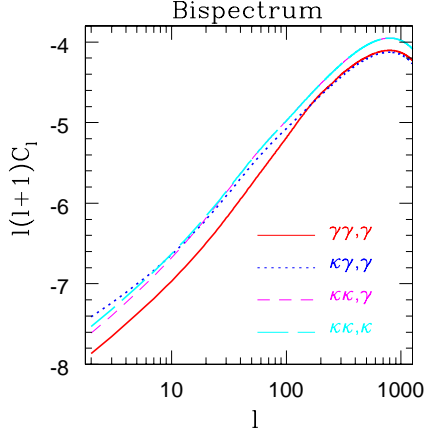}}}
\end{center}
\caption{The plots correspond to the bispectrum related power spectrum $C_l^{(2,1)}$. Various curves
correspond to the spectra $C_l^{\kappa\kappa,\kappa}$, $C_l^{\kappa\kappa,\gamma_{+}}$,
$C_l^{\kappa\gamma_{+},\gamma_{+}}$ and $C_l^{\gamma_{+}\gamma_{+},\gamma_{+}}$ as depicted. 
The source redshift is $z_1=z_2=1$. See text for details.}
\label{fig:bi}
\end{figure}

\subsection{Power spectrum associated with the Bispectrum}

The bispectrum is the lowest in the hierarchy of multispectra and have the highest Signal-to-Noise.
There has been lot of interest in modeling and detection of weak lensing bispectrum mainly using
convergence maps as well as shear. As remarked earlier, shear correlation functions are more complicated
and have rich patterns given their spinorial nature. They have been studied in real space for
specific triplet configurations analytically \citep{BerVanMell02} and have been measured in real surveys \citep{BerMellWaer03}
for first detection of non-Gaussianity signal from weak lensing observations. \cite{Ber05} has also
provided analytical results for more general configurations but in an idealistic one-halo configuration.
In this paper we focus on the power spectra associated with bispectra to probe higher order
correlation hierarchy in studying gravity induced non-Gaussianity. Our results are completely
general and given any models of underlying bispectrum can predict observed power spectrum related to
bispectrum in the presence of arbitrary observational mask. We will assume noise to be Gaussin
through out and will not contribute to the bispectrum.

\subsubsection{The Two-to-One Power spectrum: $C_l^{(2,1)}$}

We will start by considering two fields on a surface of a sphere ${\gg}(\oh)$ and ${\ggp}(\oh)$,
which are respectively of spin $s$ and $s'$. These objects can be the shear,
$\gamma_{\pm}$, flexions, $\cal F$, $\cal G$  or a tracer field $\Psi$.  The resulting product fields such as e.g.
$\gamma_{+}\gamma_{-}$ can be of
spin zero, on the other hand $\gamma_{+}^2$ and $\gamma_{-}^2$ are of spin $-4$ and $+4$ respectively. The
results can also involve fields such as $\Psi\gamma_{+}$ which is
a spin $-2$ field. A spin $s$ field can be decomposed using spin-harmonics ${}_sY_{lm}(\oh)$.
These spin harmonics are generalization of ordinary spherical harmonics $Y_{lm}(\oh)$ which are
used to decompose the spin-$0$ (or scalar) functions e.g. $\Psi(\oh)$ defined over a surface of
a sphere (sky). Throughout we will be using lower case
symbols $s,s'$ to denote the spins associated with the corresponding fields denoted in italics.
The fields that are constructed from various powers of $\gamma_{+}$ and $\gamma_{-}$ can be expanded
in terms of the corresponding spin-harmonics basis ${}_sY_{lm}(\oh)$.
The product field such as  $[{\gg}(\br){\ggp}(\br)]$ which is of spin $s+s'$ can therefore be
expanded in terms of the harmonics
${}_{s+s'}Y_{lm}(\oh)$. The following relationship therefore expresses the harmonics of
the product field in terms of the harmonics of individual fields \footnote{It is possible to prefix the
spin $s$ objects as ${}_s\sg(\br)$, however to simplify notations we will associate spin $s$ with $\sg(\br)$,
$s'$ with $\sgp(\br)$ and so on, through out the paper, and the spin $s$ should be obvious from the context.}.

\beqa
&& [{\gg}({\bf r}){\ggp}({\bf r})]_{lm} = \int d\oh {\gg}({\bf r}){\ggp}({\bf r})[{}_{s+s'}Y_{lm}^*(\oh)]
= \sum_{l_im_i} \gg_{l_1m_1}(r) \ggp_{l_2m_2}(r) \int [{}_sY_{l_1m_1}(\oh)][{}_{s'}Y_{l_2m_2}(\oh)]
[{}_{s+s'}Y_{lm}^*(\oh)] d\oh; \nn \\
&&  = \sum_{l_im_i}\gg_{l_1m_1}(r) \ggp_{l_2m_2}(r) I_{l_1l_2l}\left ( \begin{array} { c c c }
     l_1 & l_2 & l \\
     s & s' & -(s+s')
  \end{array} \right )\left ( \begin{array} { c c c }
     l_1 & l_2 & l \\
     m_1 & m_2 & m
  \end{array} \right );
\quad I_{l_1l_2l} = \sqrt {(2l_1+1)(2l_2+1)(2l+1) \over 4\pi }
\label{eq:decompose}
\eeqa

\n
The expression derived above is valid for all-sky coverage. Mixing of Electric-$E$ and Magnetic-$B$ modes results from
partial sky coverage. The results will be generalized
later to take into account arbitrary partial sky coverage and we will derive exact form for the mixing matrix.
We have assumed a Gaussian noise resulting from intrinsic ellipticity distribution of galaxies, however this
will not contribute to this non-Gaussianity statistic. We define the associated power spectra to write:

\beqa
&& C_l^{\sg \sgp, \sgpp}(r_1,r_2) = {1 \over 2l+1} \sum_m [\gg\ggp]_{lm}(r) \ggpp^*_{lm}(r) \\
&& \qquad\qquad = \sum_{l_1l_2} J_{l_1l_2l}~B^{\sg\sgp\sgp}_{l_1l_2l}(r_1,r_1,r_2)  
\left ( \begin{array} { c c c }
     l_1 & l_2 & l \\
     s & s' & -(s+s')
  \end{array} \right ); 
\qquad J_{l_1l_2l} = \sqrt {(2l_1+1)(2l_2+1) \over ( 2l+1 )}.
\label{eq:bispec}
\eeqa

\n
The power spectrum $C_l^{\sg \sgp, \sgpp}(r_1,r_2)$ essentially cross-correlates the product maps at a given radial
distance $r_1$ to another map  at another redshift $r_2$ to probe the third order statistics bispectrum $B$.
Similar statistics in the coordinate space has been reported before. \citep{berludoMell03} studied 
$\langle \gamma^2(\oh)\gamma(\oh') \rangle$
which directly deals with shear maps as opposed to convergence maps. This statistics along with a similar but simpler
version which uses $\langle\kappa^2(\oh)\kappa(\oh')\rangle$ was studied. They employed perturbation theory to
model underlying mass distribution and used a flat sky approximation to simplify their calculations.
A complementary statistics $\langle (\gamma(\oh_1)\cdot\gamma(\oh_2))\gamma \rangle$ was also considered
which relies on more detailed modeling of the bispectrum. These statistics were used by \citep{BerVanMell02}
later to detect non-Gaussianity from  VIRMOS-DESCART Lensing Survey.

Our results presented here deal with power spectrum
associated with the higher order multispectra and are derived using generic all-sky treatment and
can handle decomposition into Electric and Magnetic components in a much more straightforward manner.
Our results are complementary to their analysis. We will also generalize these results to higher order and
show how to take into account the mask in a generic way. The results presented here are not only applicable
to shear or convergence but are also applicable for higher order spinorials such as {\it Flexions}.

We have introduced the bispectrum  $B^{\sg\sgp\sgpp}_{l_1l_2l}(r_1,r_2,r_3)$ in Eq.(\ref{eq:bispec}) which
can now be related to the harmonics of the relevant fields by the following equation:

\beq
B_{l_1l_2l_3}^{\sg\sgp\sgpp}(r_1,r_2,r_3) = \sum_{m_1m_2m_3} \langle\gg_{l_1m_1}(r_1) \ggp_{l_2m_2}(r_2) \ggpp_{l_3m_3}(r_3) \rangle_c
\left ( \begin{array} { c c c }
     l_1 & l_2 & l_3 \\
     m_1 & m_2 & m_3
  \end{array} \right ).
\eeq

The matrices denote Wigner 3j symbols \citep{Ed68} which are only non zero when the quantum numbers
$l_i$ and $m_i$ satisfy certain conditions. For $x=y=0$ this generalizes the results
obtained in \citep{Cooray01} valid for the case of scalars (see also \cite{ChenSzapudi} for related discussions).
The result derived above is valid for a general $E$ and $B$ type polarization field.
Contribution from ($B$-type) magnetic polarization is believed to be considerably smaller compared to the ($E$-type) electric polarization.
The results derived above will simplify considerably if we ignore the $B$ type polarization field in our analysis. In general the power
spectra described above will not be real though real and imaginary parts can be separated by considering
different components of the bispectrum. However if we assume that the magnetic part of the polarization
is zero the following equalities will hold:

\beq
B_{l_1l_2l_3}^{\gamma_{\pm}\Psi \Psi}  =  F_{l_1} B_{l_1l_2l_3}^{\kappa\Psi\Psi}; ~~~~ \\
B_{l_1l_2l_3}^{\gamma_{\pm}\gamma_{\pm}\Psi}  = F_{l_1}F_{l_2}B_{l_1l_2l_3}^{\kappa\kappa\Psi}; ~~~\\
B_{l_1l_2l_3}^{\gamma_{\pm}\gamma_{\pm}\gamma_{\pm}}  =  F_{l_1}F_{l_2}F_{l_3}B_{l_1l_2l_3}^{\kappa\kappa\kappa}.  ~~~ \\
\eeq

The result presented above is valid for all-sky coverage. Clearly
from practical considerations we need to add a foreground mask. If we consider the
masked harmonics associated with the mask, for a given $w(\oh)$ arbitrary mask , then
expanding the product field, in the presence of mask, we can write:

\beqa
&& [{\gg}({\bf r}){\ggp}({\bf r})w(\oh)]_{lm} =
\sum_{l_im_i;l_am_a}  \gg_{l_1m_1}(r) \ggp_{l_2m_2}(r) w_{l_am_a} \int [{}_sY_{l_1m_1}(\oh)][{}_{s'}Y_{l_2m_2}(\oh)] [Y_{l_am_a}]
[{}_{s+s'}Y_{lm}^*] d\oh \nn \\
&& = \sum_{l_im_i}\sum_{l_am_a} (-1)^{l'}\gg_{l_1m_1}(r) \ggp_{l_2m_2}(r) w_{l_am_a} I_{l_1l_2l} \left (  \begin{array} { c c c }
     l_1 & l_2 & l' \\
     s & s' & -(s+s')
  \end{array} \right ) \left (\begin{array} { c c c }
     l_1 & l_2 & l' \\
     m_1 & m_2 & -m'
  \end{array} \right ) \int [{}_{-(s+s')}Y_{l'm'}^*] Y_{l_am_a}[{}_{s+s'}Y_{lm}^*] \nn \\
&& \qquad \qquad \qquad\qquad = \sum_{l_im_i} (-1)^{l+l'}\sum_{l_am_a} \gg_{l_1m_1}(r) \ggp_{l_2m_2}(r) w_{l_am_a} I_{l_1l_2l'}I_{l'l_al}
  \left (  \begin{array} { c c c }
     l_1 & l_2 & l' \\
     s & s' & -(s+s')
  \end{array} \right ) \left (\begin{array} { c c c }
     l_1 & l_2 & l' \\
     m_1 & m_2 & -m'
  \end{array} \right ) \nn \\
&& \qquad \qquad \qquad\qquad \qquad \qquad\qquad \qquad \qquad\qquad \qquad \qquad
 \times \left (  \begin{array} { c c c }
     l' & l_a & l \\
     (s+s') & 0 &  -(s+s')
  \end{array} \right ) \left (\begin{array} { c c c }
     l' & l_a & l \\
     m' & m_a & -m
  \end{array} \right ).
\eeqa

In simplifying the relations derived in this section we have used the relationship Eq.(\ref{eq:harmonics1})
and Eq.(\ref{eq:harmonics2}). We have also used the fact that ${}_sY_{lm}^* = (-1)^{m+s}Y_{l,-m}$,
where ${}^*$ denotes the complex conjugate. Similarly we can express the pseudo-harmonics of the
field $\calZ(\oh)$ observed with the same mask in terms of its all-sky harmonics.

\beqa
[{\ggpp}w]_{lm}(r) &=& \int d\oh [{\ggpp}(\oh)w(\oh)] [{}_{s''}Y_{lm}^*(\oh)]
=\sum_{l_im_i} \ggpp_{l_3m_3}(r) w_{l_bm_b}  \int [{}_{s''}Y_{l_3m_3}(\oh)] [Y_{l_bm_b}(\oh)]
[{}_{s''}Y_{lm}^*(\oh)] d\oh \nn \\
&& = \sum_{l_3m_3}\sum_{l_bm_b} \ggpp_{l_2m_2}(r) w_{l_bm_b} I_{l_3l_bl} \left (  \begin{array} { c c c }
     l_3 & l_b & l \\
     s'' & 0 & -s''
  \end{array} \right ) \left (\begin{array} { c c c }
     l_3 & l_b & l \\
     m_3 & m_b & -m
  \end{array} \right ).
\eeqa

The harmonics of the composite field $[{\sg}(\oh){\sgp}(\oh)]$ when constructed in a
partial sky are also a function of the harmonics of the mask used $w_{lm}$. The simplest
example of a mask would be $w=1$ within the observed part of the sky and $w=0$ outside.
More complicated mask
The harmonics  $w_{lm}$ are constructed out of spherical harmonics transforms.
We also need to apply the mask to the third field which we will be using in our
construction of bispectrum related power spectrum or skew-spectrum associated
with these three fields. The masks in each case are the same, however the results
could be very easily generalized for two different masks.
The {\it pseudo} power spectrum $\tilde C_l$
is constructed from the masked harmonics of relevant fields which is a cross-correlation
power spectrum:

\beqa
\tilde C_l^{\sg\sgp,\sgp}(r_1,r_2) = {1 \over 2l+ 1}
\sum_{m=-l}^{l} [{\gg}(\oh){\ggp}(\oh)w(\oh)]_{lm}(r_1) [{\ggpp}(\oh)w(\oh)]^*_{lm}(r_2).
\eeqa

\n
The pseudo power spectrum $\tilde C_l$ is a linear combination of its
all-sky counterpart $C_l$. In this sense masking introduces a coupling of modes of various order
which is absent in case of all-sky coverage. The matrix $M_{ll'}$ encodes the information regarding
the mode-mode coupling and depends on power spectrum of the mask $w_l$. For surveys with fractional sky coverage the matrix is not
invertible which signifies the loss of information due to masking. Therefore a binning of the pseudo skew- or kurt-spectrum may
be necessary before it can be inverted which leads to recovery of (unbiased) binned spectrum.

\beq
\tilde C_l^{\sg\sgp,\sgpp}(r_1,r_2) = \sum_{l'} M^{ss',s''}_{ll'} C_{l'}^{\sg\sgp,\sgpp}(r_1,r_2)
\eeq

\n
The all-sky power spectrum $C_l^{\sg\sgp,\sgpp}$ can be recovered by inverting the equation
which is related to the accompanying bispectrum by the Eq.(\ref{eq:bispec}) which
we discussed before. The mode-mode coupling matrix depends not only on the power
spectrum $|w_{l}|$ of the mask but
also of the spin associated with various fields which are being probed in construction of
skew-spectrum or the power-spectrum related to the bispectrum.

\beq
M_{ll'}^{ss',s''} = {1 \over 4\pi }\sum_{l_a} (2l'+1)(2l_a+1) \left ( \begin{array}{ c c c }
     l & l_a & l' \\
     s+s' & 0 & -{(s+s')}
  \end{array} \right )
\left ( \begin{array} { c c c }
     l & l_a & l' \\
     s'' & 0 & -s''
  \end{array} \right ) |w_{l_a}|^2
\eeq

The expression reduces to that of temperature bispectrum if we set all the spins to be zero $x=y=z=0$
in which case the coupling of various modes due to partial sky coverage only depends on the
power spectrum of the mask. In case of temperature (spin-$0$) analysis we have seen that the
mode-mode coupling matrix do not depend on the order of the statistics. The {\it skew-spectrum}
or the power spectrum related to bispectrum, as well as its higher order counterparts such as
power spectrum related to tri-spectrum can all have the same mode-mode coupling matrix
in the presence of partial sky coverage. However this is not the case for PS related to
the polarization multispectra. It depends on the spin of various fields used to construct the
bispectrum. It is customary to define a single number associated with each of these bispectrum.
The {\it skewness} is a weighted sum of the power spectrum related to the bispectrum
$S_3^{\sg\sgp,\sgpp} = \sum_l (2l+1) C_l^{\sg\sgp,\sgpp}$.

We list the specific cases of interest below. The relations between the bispectra and the
associated power spectra generalizes previously obtained cases where only the temperature
bispectrum was considered \citep{Cooray01}. These results were obtained in the context of
cosmic microwave background (CMB) polarization studies recently \citep{Mu3D10}. The radial dependence
in case of weak lensing observables needs to be properly taken into account.

\beqa
&& C_l^{\Psi\Psi,\gamma_{\pm}}(r_1,r_2) =
\sum_{l_1l_2} J_{l_1l_2l} B^{\Psi\Psi E}_{l_1l_2l}(r_1,r_1,r_2) 
\left ( \begin{array} { c c c }
     l_1 & l_2 & l \\
     0 & 0 & 0
  \end{array} \right ) \\
&& M_{ll'}^{00,2} = {1 \over 4\pi }\sum_{l_a} (2l'+1)(2l_a+1) \left ( \begin{array}{ c c c }
     l & l_a & l' \\
     0 & 0 &  0
  \end{array} \right )
\left ( \begin{array} { c c c }
     l & l_a & l' \\
     2 & 0 & -2
  \end{array} \right ) |w_{l_a}|^2.
\label{eq:TTE}
\eeqa

\n
Next we can express the power spectrum $C_l^{\Phi E,E}$ probing the mixed bispectrum $B^{\Phi EE}_{l_1l_2l}$
by the following relation:

\beqa
&& C_l^{\gamma_{\pm}\psi,\gamma_{\pm}}(r_1,r_2) =
\sum_{l_1l_2} J_{l_1l_2l} B^{\Psi EE}_{l_1l_2l}(r_1,r_1,r_2) 
\left ( \begin{array} { c c c }
     l_1 & l_2 & l \\
     \pm 2 & 0 & \mp 2
  \end{array} \right ) \\
&& M_{ll'}^{20,2} = {1 \over 4\pi }\sum_{l_a} (2l'+1)(2l_a+1) \left ( \begin{array}{ c c c }
     l & l_a & l' \\
     \pm 2 & 0 &  \mp 2
  \end{array} \right )
\left ( \begin{array} { c c c }
     l & l_a & l' \\
     \pm 2 & 0 & \mp 2
  \end{array} \right ) |w_{l_a}|^2.
\label{eq:TEE1}
\eeqa

\n
Similarly for the case  $C_l^{\gamma_{\pm}\gamma_{\pm},\gamma_{\pm}}$ which probes the bispectrum  $B^{EEE}_{l_1l_2l}$
`we can have the following expressions.

\beqa
&& C_l^{\gamma_{\pm}\gamma_{\pm},\gamma_{\pm}}(r_1,r_2) =
\sum_{l_1l_2} J_{l_1l_2l} B^{EEE}_{l_1l_2l}(r_1,r_1,r_2) 
\left ( \begin{array} { c c c }
     l_1 & l_2 & l \\
     \pm 2 & \pm 2 & \mp 4
  \end{array} \right ) \\
&& M_{ll'}^{22,2} = {1 \over 4\pi }\sum_{l_a} (2l'+1)(2l_a+1) \left ( \begin{array}{ c c c }
     l & l_a & l' \\
     \pm 4 & 0 &  \mp 4
  \end{array} \right )
\left ( \begin{array} { c c c }
     l & l_a & l' \\
     \pm 2  & 0 & \mp 2
  \end{array} \right ) |w_{l_a}|^2.
\label{eq:TEE2}
\eeqa

Other power spectra such as, $C_l^{\gamma_{\pm}\gamma_{\pm},\gamma_{\mp}}$ and  
$C_l^{\gamma_{\pm}\gamma_{\pm},\Psi}$ can also be dealt with
in a similar manner. The simpler case of convergence, i.e. $C_l^{\kappa\kappa,\kappa}$ which being a spin $0$ field can now
be analyzed as a special case and will reduce to the results previously obtained by  \citep{Cooray01}. 
The above results display expressions involving shear $\gamma$ and $\kappa$, similar results 
for flexions $\cal F$ and $\cal G$ can be derived using the same general result Eq.(\ref{eq:bispec}). We list
below the results involving only $\cal F$ or $\cal G$ flexions. It is of course possible 
to consider mixed bispectrum involving different combinations of $\cal F$ and $\cal G$. The
bispectra $B^{{\cal F}{\cal F}{\cal F}}$ and $B^{{\cal G}{\cal G}{\cal G}}$ given below can be 
expressed in terms of convergence bispectra using the expression Eq.(\ref{eq:flex}), i.e. we have
$B^{{\cal F}{\cal F}{\cal F}}_{l_1l_2l_3} = 
F^{\cal F}_{l_1}F^{\cal F}_{l_2}F^{\cal F}_{l_3}B^{{\kappa}{\kappa}{\kappa}}_{l_1l_2l_3}$ with $F^{\cal F}_l$
defined similar to the case involving shear.

\beqa
&& C_l^{{\cal F}{\cal F},{\cal F}}(r_1,r_2) =
\sum_{l_1l_2} J_{l_1l_2l} B^{{\cal F}{\cal F}{\cal F}}_{l_1l_2l}(r_1,r_1,r_2) 
\left ( \begin{array} { c c c }
     l_1 & l_2 & l \\
     1 & 1 & -2
  \end{array} \right ) \\
&& M_{ll'}^{11,1} = {1 \over 4\pi }\sum_{l_a} (2l'+1)(2l_a+1) \left ( \begin{array}{ c c c }
     l & l_a & l' \\
     1 & 1 &  -2
  \end{array} \right )
\left ( \begin{array} { c c c }
     l & l_a & l' \\
     1 & 0 & -1
  \end{array} \right ) |w_{l_a}|^2.
\label{eq:FFF}
\eeqa

\n
The power spectrum related to the second flexion $\cal G$ can likewise be written in terms of its
bispectrum  $B^{{\cal G}{\cal G}{\cal G}}$.

\beqa
&& C_l^{{\cal G}{\cal G},{\cal G}}(r_1,r_2) =
\sum_{l_1l_2} J_{l_1l_2l} B^{{\cal G}{\cal G}{\cal G}}_{l_1l_2l}(r_1,r_1,r_2) 
\left ( \begin{array} { c c c }
     l_1 & l_2 & l \\
     -3 & -3 & 6
  \end{array} \right ) \\
&& M_{ll'}^{33,3} = {1 \over 4\pi }\sum_{l_a} (2l'+1)(2l_a+1) \left ( \begin{array}{ c c c }
     l & l_a & l' \\
     -3 & -3 & 6
  \end{array} \right )
\left ( \begin{array} { c c c }
     l & l_a & l' \\
     -3 & 0 & 3
  \end{array} \right ) |w_{l_a}|^2.
\label{eq:GGG}
\eeqa

 As is clear that the
coupling matrix $M$ for different cases, depends not only on the power spectrum of the mask but also on the spins associated with
the fields that are being probed. The inversion of coupling
matrix $M_{ll'}$ can require binning. The binned coupling matrix $M_{bb'}$ in case of small sky coverage
will lead to recovery of binned power spectra $C_{l_b}$. The mask is important even for
small sky coverage as the observed region can have non-trivial topology because of
presence of very bright stars or other similar foreground objects.

The different power spectra such as $C_l^{\kappa\kappa,\kappa}$ and  $C_l^{\gamma\gamma,\gamma}$ though intrinsically
probe the same bispectrum $B^{EEE}$ weights individual modes in a different manner and can be used to
probe effect of systematics in handling real data. Each of these power spectra can be used to define
corresponding one-point skewness. We plot the results of our numerical calculations in Fig-\ref{fig:bi}
for various bispectrum related power spectrum. The power spectrum for convergence for the same models
are displayed in Fig-\ref{fig:ps}. For these calculations we put all the sources at a same redshift $z_s=1$.
The results can take into account for a source redshift distribution with a given spread and median redshift
of sources.

\subsection{Power spectrum associated with the Trispecrum}

Where as the bispectrum represents the lowest order deviation from the Gaussianity, the
next higher order representation is Trispectrum. Reason to go beyond lowest order
non-Gaussianity is multifold. While lowest order non-Gaussianity indeed probed
directly by bispectrum with a higher signal-to-noise, the study of trispectrum
can go beyond lowest order in probing the gravity induced non-Gaussianity.

In a different context, relating to CMB studies involving secondaries it is known that
trispectrum related power spectra $C_l^{(2,2)}$ can separate out lensing induced non-linearity from
other contributions from secondaries without involving cross-correlations with any external data sets
see e.g.  \citep{MuCross09} for extensive discussions. It can also put independent
constraints on non-Gaussianity parameters $g_{NL}$ and $f_{NL}$. The power spectrum $C_l^{(3,1)}$
associated with the Trispectrum has also been used in studies of 21cm surveys
which maps neutral hydrogen distribution at higher redshift \citep{Cooray06,CooLiMel08} .

We explore here the possibility of using these power spectra for studies in the context of weak
lensing. To model weak lensing tri-spectra one needs to model the underlying mass trispectra.
This we do here using a well motivated model, namely hierarchical ansatz which is known
to be valid at smaller angular scales. However we want to stress that modeling of
underlying trispectra can always follow more complicated prescriptions from
(extensions of) perturbation theory or various improvements halo models which are also employed frequently
in such studies.

\subsubsection{The Two-to-Two Power spectrum: $C_l^{(2,2)}$}

In constructing the trispectrum related power spectrum we start with two fields
$\gg$ and $\ggp$ respectively on the surface of the sphere. As before we can
take specific examples where these fields are either $\gamma_{\pm}$ or $\Psi$.
We will keep the analysis generic here and will consider the specific examples later on.
The spins associated with various fields are denoted by lower case symbols, i.e. $u$ and $v$. The
product filed now can be expanded in terms of the spin harmonics of spin $u+v$.
as was done in Eq.(\ref{eq:decompose}). Similarly decomposing the other set of
product field  we obtain $[\ggpp(\oh)\ggppp(\oh)]_{lm}$. We next construct the
the power spectrum associated with the trispectrum from these harmonics:

\beq
C_l^{\sg\sgp,\sgpp\sgppp}(r_1,r_2) = {1 \over 2l +1} \sum_{m=-l}^l [\gg\ggp]_{lm}(r_1) [\ggpp\ggppp]_{lm}^*(r_2)
\eeq

\n
This particular type of power spectra associated with trispectra has been studied extensively
in the literature in the context of CMB studies; especially to separate out effects due to lensing of CMB from other
secondaries. This is one of the two degenerate power spectra
associated with trispectrum. After going through very similar algebra outlined in the previous section we can express the
$C_l^{\sg\sgp,\sgpp\sgppp}$ in terms of the relevant trispectra which it is probing. The expression in the
absence of any mask takes the following form:

\beqa
&& C_l^{\sg\sgp,\sgpp\sgppp}(r_1,r_2) = \sum_{l_1,l_2,l_3,l_4} J_{l_1l_2l}J_{l_3l_4l}~
(T^{\sg_{l_1}\sgp_{l_2}}_{\sgpp_{l_3}\sgppp_{l_4}}(L) - G^{\sg_{l_1}\sgp_{l_2}}_{\sgpp_{l_3}\sgppp_{l_4}}(L))
 \left ( \begin{array}{ c c c }
     l_1 & l_2 & l \\
     s & s' & -{(s+s')}
  \end{array} \right )
\left ( \begin{array} { c c c }
     l_3 & l_4 & l \\
     s'' & s''' & -{(s''+s''')}
  \end{array} \right ) \nn \\
&& \qquad \qquad \qquad \qquad \qquad 
 \qquad \qquad \qquad \ggpp, \ggppp, \ggppp \in \gamma_{\pm}, \Psi, {\cal F}, {\cal G}
\eeqa

\n
The Gaussian contribution from the disconnected part of the trispectrum $G$ needs to be subtracted to construct the estimator.
In the presence of a completely general mask $w(\oh)$ the pseudo-$C_{\ell}$s or PCLs will have to be modified
to take into account the effect of mask. This involves computing the spherical harmonics of the masked field
$[\sg(\oh)\sgp(\oh)w(\oh)]$ and cross-correlating it against the harmonics of $[\sgpp(\oh)\sgpp(\oh)w(\oh)]$.

\beq
\tilde C_l^{\sg\sgp,\sgpp\sgppp}(r_1,r_2) = {1 \over 2l +1} \sum_{m=-l}^l [\gg\ggp~w]_{lm} [\ggpp\ggppp~w]_{lm}^*;
\qquad \qquad w_{l} = {1 \over 2l+1} \sum_{m=-l}^l w_{lm}w_{lm}^*
\eeq

The resulting PCLs are linear combination of their all-sky counterpart. The mixing matrix
which encode the information about the mode mixing will depend on the power spectrum of the
mask as well as the spins of all four associated fields. The mixing matrix $M_{ll'}$
expressed in terms of the Wigner's $3j$ symbols take the following form:

\beq
M_{ll'}^{ss',s''s'''} = {1 \over 4\pi }\sum_{l_a} (2l'+1)(2l_a+1) \left ( \begin{array}{ c c c }
     l & l_a & l' \\
     s+s' & 0 & -{(s+s')}
  \end{array} \right )
\left ( \begin{array} { c c c }
     l & l_a & l' \\
     s''+s''' & 0 & -(s''+s''')
  \end{array} \right ) |w_{l_a}|^2; \qquad \qquad s,s',s'',s''' \in0,1, \pm2,3;
\eeq

\n
The spin indices $s,s'$,\dots take values $0$ for $\kappa$, $\pm2$ for $\gamma_{\pm}$,$1$ for {\cal F} and
$3$ for {\cal G}. The pseudo-$C_\ell$s expressed as a linear combination of all-sky power spectra can now be expressed
using the following relationship:

\beq
\tilde C_l^{\sg\sgp,\sgpp\sgppp}(r_1,r_2) = \sum_{l'} M_{ll'}^{ss's'',s'''} C_{l'}^{\sg\sgp,\sgpp\sgppp}(r_1,r_2)
\eeq

For near all sky surveys and with proper binning the mixing matrix $M_{ll'}$ can be made invertible. This
provides an unique way to estimate all-sky  $C_{l'}^{UVW,X}$ and the associated information contents
about the trispectra. For a given theoretical prediction the all-sky power spectra $C_{l'}^{\sg\sgp\sgpp,\sgppp}$
can be analytically computed. Knowing the detailed model of an experimental mask allows us to
compute the observed $\tilde C_l^{\sg\sgp\sgpp,\sgppp}$ accurately. The results presented here generalizes the
ones obtained in \citep{Munshi_kurt,MuPol10} for the case of shear and Flexions.
While the shear $\gamma$ like the polarization fields $Q \pm iU$ are spin -$2$ objects, flexions on the
other hand can be associated with higher (or lower) spins.

These results assumes generic field variables $\gg({\bf r})$, ${\ggp}({\bf r})$ which can have arbitrary
spin associated to them. We next specifiy certain specific cases where we identify three of the fields
${\gg, \ggp, \ggpp} = \Psi$ and ${\ggppp}= \gamma_{\pm}$. The other combinations can also be obtained in a similar manner.

\beqa
&& C_l^{\Psi\Psi,\Psi \gamma_{\pm}} = \sum_{l_1,l_2,l_3,l_4}
J_{l_1l_2l}J_{l_3l_4l}~T^{\Psi_{l_1}\Psi_{l_2}}_{\Psi_{l_3}E_{l_4}}(l)
\left ( \begin{array}{ c c c }
     l_1 & l_2 & l \\
     0 & 0 & 0
  \end{array} \right )
\left ( \begin{array} { c c c }
     l_3 & l_4 & l \\
     \pm 2 & 0 & \mp 2
  \end{array} \right ); \\
&& M_{ll'}^{00,02} = {1 \over 4\pi }\sum_{l_a} (2l'+1)(2l_a+1) \left ( \begin{array}{ c c c }
     l & l_a & l' \\
     0 & 0 &  0
  \end{array} \right )
\left ( \begin{array} { c c c }
     l & l_a & l' \\
     \pm 2 & 0 & \mp 2
  \end{array} \right ) |w_{l_a}|^2;
\eeqa

\n
Other estimators for mixed trispectra involving different combinations of $E$-polarization and a tracer field (scalar) $\Psi$ can be
derived in a similar manner and can provide independent information of corresponding trispectra. We have ignored the presence of
$B$-type polarization in our analysis. Presence of non-zero B-mode can be dealt with very easily in our framework but the resulting
expressions will be more complicated.

\subsubsection{The Three-to-One Power spectrum: $C_l^{(3,1)}$}

The number of power spectra that can be associated with a given multispectra depends on number of different way
the order of the multispectra can be decomposed into a pair of integers. The bispectrum being of order
three can be decomposed uniquely $3=2+1$ and has only one associated power spectra. On the other hand
the trispectrum whose order ($=4$) can be decomposed in a two different way; i.e., $4=3+1=2+2$. Hence trispectrum of a
specific type correspond to a pair of two different power spectrum associated with it (see e.g. \cite{Munshi_kurt} for more details).
The results presented here are generalization for the case of non-zero spins.

The other power spectrum associated with the trispectrum is constructed by cross-correlating product of
three different fields $[\gg\ggp\ggpp]$ with the remaining field $\ggppp$. The cross correlation
power spectrum in terms of the multipoles are given by the following expression:

\beqa
&& [\gg(\br)\ggp(\br)\ggpp(\br)]_{lm} = \int [\gg\ggp\ggpp][{}_{s+s'+s''} Y_{lm}(\oh)^*]d\oh; \qquad
[\ggppp(\br)]_{lm} = \int [\ggppp][{}_{s'''} Y_{lm}(\oh)^*]d\oh; \qquad \\
&& ~~~~~~~~~~~~~~~~~~ C_l^{\gg\ggp\ggpp,\ggppp}(r_1,r_2) = {1 \over 2l + 1} \sum_{m=-l}^l [\gg\ggp\ggpp]_{lm}(r_1) [\ggppp]_{lm}^*(r_2).
\eeqa

\n
By repeated use of the expressions Eq.(\ref{eq:harmonics1}) or equivalently Eq.(\ref{eq:harmonics2}) to simplify the
harmonics of the product field in terms of the individual harmonics, we can express the all-sky result in the following
form:

\beqa
C_l^{\sg\sgp\sgpp,\sgppp}(r_1,r_2) = && \sum_{l_1,l_2,l_3,L} J_{l_1l_2L}J_{Ll_3l}~ T^{\sg_{l_1}\sgp_{l_2}}_{\sgpp_{l_3}\sgppp_l}(L)
\left ( \begin{array}{ c c c }
     l_1 & l_2 & L \\
     s & s' & -{(s+s')}
  \end{array} \right )
\left ( \begin{array} { c c c }
     L & l_3 & l' \\
     (s+s') & s'' & -{(s+s'+s'')}
  \end{array} \right ) \nn  \\
&& \qquad\qquad \qquad 
\qquad \qquad s,s',s'',s''' \in0,1, \pm2,3.
\eeqa

\n
In case of of partial sky coverage with a generic mask $w(\oh)$ the relevant expression for the Pseudo-$C_{\ell}$s
will involve the harmonic transform of the mask.

\beqa
&& [\gg(\br)\ggp(\br)\ggpp(\br) w(\oh)]_{lm} = \int [\gg\ggp\ggpp w][{}_{u+v+w} Y_{lm}^*(\oh)]d\oh; \qquad \nn \\
&& \qquad \qquad [\ggppp(\br)w(\oh)]_{lm} = \int [\ggppp w][{}_{s'''} Y_{lm}^*(\oh)]d\oh; \qquad \tilde C_l^{\gg\ggp\ggpp,\ggppp} =
 M_{ll'}C_{l'}^{\gg\ggp\ggpp,\ggppp}
\eeqa

\n
The mixing matrix has the following expression in terms of various spins involved and the power spectra of the
mask introduced before. Note that the mixing matrix is of different form compared to what we obtained for
two-to-one power spectra. This is related to how various fields with different spins were combined to
construct these two estimators. In case of convergence trispectrum (spin-0) the two mixing matrices
take the same form.

\beq
M_{ll'}^{ss's'',s'''} = {1 \over 4\pi }\sum_{l_a} (2l'+1)(2l_a+1) \left ( \begin{array}{ c c c }
     l & l_a & l' \\
     (s+s'+s'') & 0 & -{(s+s'+s'')}
  \end{array} \right )
\left ( \begin{array} { c c c }
     l & l_a & l' \\
     s''' & 0 & -s'''
  \end{array} \right ) |w_{l_a}|^2;
\eeq

The expressions derived here are derived for general mask but Gaussian noise of intrinsic ellipticity distribution of Galaxies.
Any residual non-Gaussianity from systematics will have to be subtracted out. The estimators derived above indeed are unbiased but unoptimized.
Optimization of these estimators will however have to be done in a model dependent way and from pre-constructed $E$ and $B$ maps from
observed shear data.

For a specific example we choose $\gg = \ggp = \ggpp = \Psi$ and $\ggppp= E$. In this case the three-to-one
estimator takes the following form:

\beqa
&& C_l^{\Psi\Psi\Psi,\gamma_{\pm}} = \sum_{l_1,l_2,l_3,L} J_{l_1l_2L}J_{Ll_3l} ~ T^{\Psi_{l_1}\Psi_{l_2}}_{\Psi_{l_3}E_l}(L)
\left ( \begin{array}{ c c c }
     l_1 & l_2 & L \\
     0 & 0 & 0
  \end{array} \right )
\left ( \begin{array} { c c c }
     L & l_3 & l' \\

     0 & 0 & 0
  \end{array} \right )
\nn \\
&& M_{ll'}^{000,2} = {1 \over 4\pi }\sum_{l_a} (2l'+1)(2l_a+1) \left ( \begin{array}{ c c c }
     l & l_a & l' \\
     0 & 0 &  0
  \end{array} \right )
\left ( \begin{array} { c c c }
     l & l_a & l' \\
     \pm 2 & 0 & \mp 2
  \end{array} \right ) |w_{l_a}|^2;
\eeqa

\n
The other possibilities include the pure trispectrum that corresponds only to $E$-type polarization which can
be used directly from a cubic shear map cross-correlated against a shear map:

\beqa
&& C_l^{\gamma_{\pm}\gamma_{\pm}\gamma_{\pm},\gamma_{\pm}} = \sum_{l_1,l_2,l_3,L} ~J_{l_1l_2L}J_{Ll_3l} ~ 
T^{E_{l_1}E_{l_2}}_{E_{l_3}E_l}(L)
\left ( \begin{array}{ c c c }
     l_1 & l_2 & L \\
     \pm 2 & \pm 2 & \mp 4
  \end{array} \right )
\left ( \begin{array} { c c c }
     L & l_3 & l' \\
     \mp 4 & 0 & \pm 4
  \end{array} \right )
\nn \\
&& M_{ll'}^{222,2} = {1 \over 4\pi }\sum_{l_a} (2l'+1)(2l_a+1) \left ( \begin{array}{ c c c }
     l & l_a & l' \\
     \pm 6 & 0 &  \mp 6
  \end{array} \right )
\left ( \begin{array} { c c c }
     l & l_a & l' \\
     \pm 2 & 0 & \mp 2
  \end{array} \right ) |w_{l_a}|^2;
\eeqa

\n
The generalization to trispectra involving the flexions $\cal F$ and $\cal G$ will
follow a procedure similar to that of bispectrum.

The bispectrum is defined through a triangular configuration in the multipole space.
The trispectrum on the other hand is associated with a quadrangle in the multiple space.
The extraction of information by the associated power spectra defined above corresponds
to summing over all possible configurations keeping one of the sides of the triangle fixed.
In case of power spectra associated with trispectra there are two different options:
to keep the diagonal fixed and sum over all possible configurations $C_l^{(2,2)}$ or
to keep one of the sides fixed and sum over all possible configurations  $C_l^{(3,1)}$.

It is possible to however introduce a window optimized to selectively search for information for a
specific configuration either for bispectrum or for trispectrum. However such analysis which introduces
mode-mode coupling can not be generalized to the arbitrary partial sky coverage as there is already a coupling of modes
because of non-uniform coverage of the sky.

Unlike the bispectra, the trispectra has non-vanishing contribution from the Gaussian component.
The Gaussian contribution represents the unconnected component of the total Trispectra and needs to be
subtracted out. These can be treated by using a Monte-Carlo approach which uses identical mask.
In our analysis we have used the harmonic space approach which is complementary to
real space analysis. It is possible to device statistics such as collapsed four-point functions,
$\langle \gamma^2(\oh)\gamma^2(\oh') \rangle_c$ or $\langle \gamma^3(\oh)\gamma(\oh)\rangle_c $
which will generalize the bispectrum related
statistics proposed by \citep{BerVanMell02,berludoMell03} in real space, for the case of three-point correlation functions,
to four-point correlation function in real space or equivalently to power spectrum related to trispectra.
However higher order correlation function as well as their Fourier transforms will be more
dominated by noise and higher number density of galaxies will be required to probe
non-Gaussianity beyond the bispectrum.

We also want to point out that at higher redshift, the number density of sources will be low.
This will result in  increase of shot noise. It is possible to include a redshift dependent
weight function $w(r)$. In the simplest case, this would reflect the number density of source
galaxies. Such a weighting scheme can be incorporated in the modeling of our mask, which we have
left completely arbitrary. If we use two different set of mask for two different redshift bands
we can still use the expressions derived above for coupling matrices $M_{ll'}$. However we need to replace
the power spectrum $|w_l|^2$ of the mask with a cross-spectra of two different masks that are being used.

It is possible to compute the covariance of our estimates of various $C_l$s, i.e. $C_lC_{l'}$ under
certain simplifying assumptions which involves modeling of higher order correlations accurately.
The shot noise contributions, involving various powers of source galaxy density, will dominate
at higher $l$. A detailed analysis will be presented elsewhere.

\section{Modelling Gravity induced Non-Gaussianity}

We start by modeling the bispectrum and trispectrum of the convergence field $\kappa(\oh)$ in terms
of the underlying mass distribution $\delta$. Next we will link the bispectrum and trispectrum
 associated with the shear $\gamma(\oh)$ or flexions ${\cal F}(\oh)$ or ${\cal G}(\oh)$ fields in terms
of the corresponding quantities for the convergence $\gamma$. These will be used for the computation
of the power spectra  associated with these multispectra of various order.

\subsection{Linking Weak Lensing Observables and Mass Correlation Hierarchy}

The two-point and higher order correlation functions as well as their Fourier
counterparts, the power spectrum, the bispectrum and other higher order multispectra
beyond the lowest orders are well understood for weak lensing convergence
field \citep{MuJa00,MuJai01}. The modeling depends on accurate modeling of underlying mass distribution.
The projected convergence power spectrum $C_l^{\kappa}$ can be expressed in terms of the 3D mass power spectrum $P^\delta$:

\be
C_l^{\kappa} = \int dr {w^2(r) \over d_A^2} P^\delta \left ( {l \over d_A(r)};r \right ); \qquad
w(r) = {3 \Omega_M H_0^2 \over 2 a c^2 }{d_A(r) d_A(r_s-r) \over d_A(r_s)}.
\label{ps}
\ee

\n
Here $d_A(r)$ is the
angular diameter distance at a comoving distance $r$, corresponding scale factor is $a$
and $r_s$ is the distance to the background source distribution (we assume that all sources
are situated at a redshift of $z_s=1$). The power spectrum $C_l^{\kappa}$ of the convergence
field $\kappa$ is therefore a weighted projection
of the 3D matter power spectrum along the line of sight. The weight $w(r)$ depend on
underlying cosmology through the Hubble constant $H_0$, $\Omega_M$ as well as
$d_A$. It also depends on the source redshift.

\be
\langle \kappa_{l_1m_1}\kappa_{l_2m_2}\kappa_{l_3m_3}\rangle_c = \left ( \begin{array} { c c c }
     l_1 & l_2 & l_3 \\
     m_1 & m_2 & m_3
  \end{array} \right ) B^{\kappa\kappa\kappa}_{l_1l_2l_3}; \qquad
B^{\kappa\kappa\kappa}_{l_1l_2l_3} = \sum_{m_1m_2m_3}\left ( \begin{array} { c c c }
     l_1 & l_2 & l_3 \\
     m_1 & m_2 & m_3
  \end{array} \right )\langle \kappa_{l_1m_1}\kappa_{l_2m_2}\kappa_{l_3m_3}\rangle_c
\ee

\n
The bispectrum of the convergence can now be expressed as the bispectrum of the
underlying mass distribution. The all-sky expression for bispectrum can be
written as:

\be
B^{\kappa\kappa\kappa}_{l_1l_2l_3} = I_{l_1l_2l_3}\int dr {w^3(r) \over d^4_A(r)} B^{\delta}
\left ( {l_1 \over d_A(r)},{l_2 \over d_A(r)},{l_3 \over d_A(r)};r \right );
\qquad I_{l_1l_2l_3}= \sqrt{(2l_1+1)(2l_2+1)(2l_3+1)\over 4\pi} \left ( \begin{array} { c c c }
     l_1 & l_2 & l_3 \\
     0 & 0 & 0
  \end{array} \right )
\label{bis}
\ee

\n
A similar expression holds for trispectrum $T^{l_1l_2}_{l_3l_4}(L)$ in terms of
matter trispectrum. The above expressions are derived within the context of Limber's approximation
as well as the Born's approximations which are widely used for such calculations.
The flat sky bispectrum $b_{l_1l_2l_3}$ is related to its all-sky counterpart
by the relation $B_{l_1l_2l_3} \equiv I_{l_1l_2l_3} b_{l_1l_2l_3}$. Using expression derived
previously Eq.(\ref{eq:flex}) or Eq.(\ref{ref:shearkappa}) we can relate the bispectrum for convergence and
the bispectrum associated with the shear or flexion field are related $B^{EEE}_{l_1l_2l_3} = 
F^E_{l_1}F^E_{l_2}F^E_{l_3} B^{\kappa\kappa\kappa}_{l_1l_2l_3}$
with $\qquad F^E_l = \sqrt {(l+2)(l-1) \over l (l+1)}$. For large values of $l$ $F_l \sim 1$ and different
power spectra that we defined associated with bispectra differ only through the geometrical
form factor as encoded in the Wigner 3j symbols.

\subsection{Matter Correlation Hierarchy}

It is clear that we need accurate analytical
modeling of dark matter clustering for prediction of weak lensing statistics. We lack detailed such theoretical
understanding of gravitational clustering. On larger scales, where the density field
is only weakly nonlinear, perturbative treatments are known to be
valid. For a phenomenological statistical description of dark matter clustering in
collapsed objects on nonlinear scales, typically
the halo model \citep{CooSeth02} is used.  We will be using the halo model
in our study. However, an alternative approach
on small scales is to employ various \emph{ansatze} which trace
their origin to field theoretic techniques used to probe
gravitational clustering. The hierarchical
\emph{ansatz} has also been used for many weak lensing related work, 
where the higher-order correlation functions are
constructed from the two-point correlation functions. Assuming a
tree model for the matter correlation hierarchy (typically used in
the highly non-linear regime) one can write the most general case,
the $N$ point correlation function, $\xi_N^\delta({\bf r}_1, \dots, {\bf
r}_n)$ as a product of two-point correlation functions
$\xi_2^\delta(|\br_i-\br_j|)$ \citep{Bernardreview02}. Equivalently in the
Fourier domain the multispectra can be written as products of the
matter power spectrum $P\ad(k_1)$. The temporal dependence is
implicit here.

\begin{equation}
\xi_N({\bf r_1}, \dots, {\bf r_n}) \equiv \langle \delta({\bf r_1})
\dots \delta({\bf r_n})\rangle_c = \sum_{\alpha,{\rm N=trees}}
Q_{N,\alpha}\sum_{\rm labellings} \displaystyle \prod_{\rm edges
(i,j)}^{(N-1)} \xi_2(|\br_i-\br_j|).
\end{equation}

\n
It is however very interesting to note that a similar hierarchy develops in the quasi-linear regime at tree-level
in the limiting case of vanishing variance. However the hierarchical amplitudes become
shape-dependent in such a case. These kernels are also used to relate the halo-halo
correlation hierarchy with underlying mass correlation hierarchy.
Nevertheless there are indications from numerical simulations that
these amplitudes become configuration-independent again as has been shown by high resolution studies
for the lowest order case $Q_3 = Q$ \citep{Scocci98, Bernardreview02}. See \cite{Waerbeke01}
for related discussion about use of perturbation theory results in intermediate scales.
In the Fourier space however such an ansatz means that the entire hierarchy of the multi-spectra can be written in terms of
sums of products of power spectra with different amplitudes $Q_{N,\alpha}$ etc. The power spectra is defined
through $\langle \delta(\bk_1)\delta(\bk_2) \rangle_c  = (2\pi)^3 \delta_{3D}({\bf k}_{12})P(k_1)$. Similarly
the bispectrum and trispectrum are defined through the following expressions
$\langle \delta(\bk_1) \delta(\bk_2)\delta(\bk_3) \rangle_c = (2\pi)^3 \delta_{3D}(\bk_{123})B^\delta(\bk_1,\bk_2,\bk_3)$ and
$\langle \delta(\bk_1) \cdots \delta(\bk_4) \rangle_c = (2\pi)^3 \delta_{3D}(\bk_{1234})T^\delta(\bk_1,\bk_2,\bk_3,\bk_4)$.
The subscript $c$ here represents the connected part of the spectra and $\bk_{i_1\dots i_n} = \bk_{i_1} + \dots + \bk_{i_n}$. The Dirac
delta functions $\delta_{3D}$ ensure the conservation of momentum at each vertex representing the multispectrum.

\begin{eqnarray}
&& B^\delta(\bk_1,\bk_2, \bk_3)_{\sum \bk_i=0} =
Q_3[P^\delta(k_1)P^\delta(k_2)+ P^\delta(k_1)P^\delta(k_3)+ P^\delta(k_2)P^\delta(k_3)] \\
&& T^\delta(\bk_1,\bk_2,\bk_3,\bk_4)_{\sum \bk_i=0} = R_a[P^\delta(k_1)P^\delta(k_2)P^\delta(k_3)
+ cyc.perm.]+ R_b[P^\delta(k_1)P^\delta(|\bk_{12}|)P^\delta(|\bk_{123}|)+ cyc.perm.].
\end{eqnarray}

\noindent
Different hierarchical models differ in the way numerical values are allotted to various amplitudes. \citet{BerSch92}
considered ``snake'', ``hybrid'' and ``star'' diagrams with differing amplitudes at various order. A new ``star''  appears at each order.
higher-order ''snakes'' or ``hybrid'' diagrams are built from lower-order ``star'' diagrams. In models where we only have only star diagrams
\citep{VaMuBa04} the expressions for the trispectrum takes the following form: $T^\delta(\bk_1,\bk_2,\bk_3,\bk_4)_{\sum \bk_i=0}  = Q_{4}[P^\delta(k_1)P^\delta(k_2)P^\delta(k_3) + cyc.perm.]$. Following \cite{VaMuBa04} we will call these models ``stellar models''.
Indeed it is also possible to use perturbative calculations which are however valid only at large scales.
While we still do not have an exact description of the non-linear clustering of a self-gravitating medium
in a cosmological scenario, theses approaches do capture some of the salient features of gravitational
clustering in the highly non-linear regime and have been tested extensively against numerical simulation in 2D
statistics of convergence of shear \citep{VaMuBa04}. These models were also used in modeling of the covariance of
lower-order cumulants \citep{MuVa05}.

\subsection{Halo Model}

Halo model on the other hand relies on modeling the clustering of halos and predictions from
perturbative calculation to model the non-linear correlation functions. It is known that the halo
over density at a given position ${\bf x}$, $\delta^h({\bf x}, M; z)$ can be related to the underlying density contrast
$\delta({\bf x}, z)$ by a Taylor expansion as was shown by \citep{Mo97}.

\be
\delta^h({\bf x}, M; z) = b_1(M;z) \delta({\bf x},z) + {1 \over 2} b_2(M,z) \delta^2({\bf x},z) + \dots
\ee

\n
The expansion coefficients are functions of the threshold $\nu_c = {\delta_c / \sigma(M,z)}$. Here $\delta_c$
is the threshold for a spherical over-density to collapse and $\sigma(M,z)$ is r.m.s fluctuation within a tophat 
filter. The halo model incorporates the perturbative aspects of gravitational dynamics by using it to
model halo-halo correlation hierarchy. The nonlinear features take direct contribution
from the halo profile. Other ingredients being number density of halo. The total power spectrum $P^t(k)$ 
at non-linear scale can be written as \citep{S00}

\be
P^{PP} = I_2^0(k,k); \qquad P^{HH}(k) = [I_1^1(k)]^2 P(k); \qquad P^t = P^{HH}(k) + P^{PP}(k).
\ee

In our calculation minimum halo mas that we consider is $10^3 M_{}$ and the maximum is $110^{16}M_{}$. It is
known that the massive halos do not consider significantly due to their low abundance. 
The bispectrum involves terms from one, two or three halo contributions and the total can be written as:

\beqa
&& B^t(k_1,k_2,k_3) = B^{PPP}(k_1,k_2,k_3)+ B^{PPP}(k_1,k_2,k_3)+ B^{PPP}(k_1,k_2,k_3); \\
&& B^{PPP} = I_3^0(k_1,k_2,k_3); \qquad B^{PHH}(k_1,k_2,k_3) = I_2^1(k_1,k_2)I_1^0(k_3)P(k_3) +cyc.perm.;\\ 
&& B^{HHH}(k_1,k_2,k_3) = [2J(k_1,k_2,k_3)I_1^1(k_3) 
+ I_1^2(k_3)] I_1^1(k_1)I_1^1(k_2)P(k_1)P(k_2)+ cyc.perm.
\label{halo_corr}
\eeqa

\n
The kernel $J(k_1,k_2,k_3)$ is derived using second order perturbation theory \cite{Fry84,Bouchet92} which we
discussed above in the context of hierarchical ansatz. The integrals $I_{\mu}^{\beta}$ can be expressed in
terms of the Fourier transform of halo profile (assumed to be an NFW \citep{NFW96} profile):

\be
I^{\beta}_{\mu}(k_1,k_2,\dots, k_\mu;z) = \int dM \left ({M \over \rho_b} \right )^\mu {dn(m,z) \over dM } b_{\beta}(M)
y(k_1,M)\dots y(k_\mu,M); \qquad y(k,M) = {1 \over M}\int_0^{r_v} dr 4\pi r^2 \rho(r,M) \left [ {\sin(kr) \over kr }\right ]
\ee
The mass function is assumed to be given by Press-Schechter mass function \citep{PS74}. Our calculations
are based on using the expressions \ref{halo_corr} in equation Eq.(\ref{ps}) and Eq.(\ref{bis}). The
convergence power spectra and bispectra are next related to the observables involving shear using 
Eq.(\ref{eq:TTE}),Eq(\ref{eq:TEE1}) and Eq.(\ref{eq:TEE2}). We will only be considering only lowest order
non-Gaussianity statistics, i.e. bispectrum here but results can readily be generalized to higher orders.

\section{Conclusions}

Future weak lensing surveys will play a crucial role in cosmological
research, particularly in further reducing the uncertainty in
fundamental properties of the standard cosmological model, including
those that describe the evolution of equation of state of dark
energy \cite{Euclid}. It is known that, by exploiting both the
angular diameter distance and the growth of structure, weak lensing
surveys are able to constrain cosmological parameters, as well as
testing the accuracy that general relativity provides as a
description of gravity \cite{HKV07,Amendola08,Benyon09,
Schrabback09, Kilbinger09}.

Constraints from weak lensing surveys are complementary to those
obtained from cosmic microwave background studies and from galaxy
surveys as they probe structure formation in the dark sector at a
relatively low redshift range. Initial studies in weak lensing were
restricted to studying two-point functions in projection for the
entire source distribution. It was, however, found that binning
sources in a few photometric redshift bins can improve the
constraints \cite{Hu99}. More recently a full 3D formalism has been
developed which uses photometric redshifts of all sources without
any binning\cite{Heav03,Castro05,HKT06}. These studies have
demonstrated that 3D lensing can provide more powerful and tighter
constraints on the dark energy equation of state parameter, on
neutrino masses \cite{deBernardis09}, as well as testing braneworld
and other alternative gravity models. Most of these 3D works have
primarily focused on power spectrum analysis, but in future accurate
higher-order statistic measurement should be possible (e.g.
\cite{TakadaJain04,Semboloni09}).

In a recent work \citep{Mu3D10} extended previous studies using
convergence maps to probe gravity-induced non-Gaussiniaty. These
studies go beyond previous analysis of power-spectrum estimation
from convergence data. Recovery of cosmological information however
is even more straightforward from shear maps. In this paper we have
generalized the results previously obtained for convergence to
shear. We have gone beyond conventional power spectrum analysis by
defining power spectra associated with higher-order multispectra in
such a way as to compress information available in the multispectra
to a single derived power spectrum. Generic results which do not
depend on detailed modeling of multispetcra have been obtained and
later specialized with concrete models using specific forms for the
gravity-induced clustering as encoded in the higher-order
multispectra. The generic results we have presented do not depend on
the  specific nature of non-Gaussianity and are able to handle
primordial as well as secondary non-Gaussianity.

Though the higher-order multispectra contain a wealth of information
in through their shape dependence, they are partly degenerate. It
would be desirable to exploit the entire information content encoded
in higher order multispectra  for constraining structure formation
scenarios. However, the direct determination of the multispectra and
their complete shape dependence from noisy data can be a very
difficult task. In this paper we have employed a set of statistics
called ``cumulant correlators'' which were first used in real space
in the context of galaxy surveys \citep{SzaSza93,SzaSza97} and later
extended to CMB studies \citep{MuPol10,MuPol09}. We have presented a
general formalism for the study of the power spectra or the Fourier
transforms of these correlators. We present a 3D analysis which
takes into account the radial as well as on the surface of the sky
decomposition. Previous studies in weak lensing mainly concentrated
on convergence maps containing spin-0 objects that can be treated
relatively easily. However, from the point of view of observational
data reduction, it is much more natural to focus on galaxy shear
which is a natural byproduct of weak lensing surveys. In this paper
we have extended the concept of cumulant correlators to higher spin
objects to {\it shear} as well as {\it flexions}. We consider the
correlations of cumulants which are constructed from such spinorial
objects. The analytical results we obtain are valid for an {\it
arbitrary} mask and our formalism allows us to correct for the
effect of mask at arbitrary order for general spinorial field. The
analytical results are sufficiently generic to include a tracer
field for the large scale structure and probe mixed bispectrum
involving underlying mass distribution as well as the large scale
tracer field.

We have restricted this study to the third and fourth order, as data
from observations gets increasingly noise dominated as the order
increases. However,  our analysis  can be generalized to higher
order and some of our results are indeed valid at arbitrary order.
At third order, we define a power spectrum which compresses
information associated with a bispectrum to a power spectrum. This
power spectrum $C_l^{\sg\sgp,\sgpp}(r_2,r_1)$ is the cross-power
spectrum associated with squared convergence maps
$\sg(r_1,\oh)\sgp(r_1,\oh)$ constructed at a specific radial
distance $r_1$ against $\sgpp(r_2,\oh)$ at $r_2$. In a similar
manner we also associate power spectra $C_l^{\sg\sgp,\sgpp\sgppp}$
and $C_l^{\sg\sgp\sgpp,\sgppp}$ with associated trispectra
$T^{l_1l_2}_{l_3l_4}(L;r_i)$. There are two different power spectra
at the level of trispectra which are related to the respective
real-space correlation functions $\langle
\sg(r_1,\oh)\sgpp(r_1,\oh)\sgpp(r_2,\oh')\sgppp(r_2,\oh)\rangle$ and
$\langle
\sg(r_1,\oh)\sgp(r_1,\oh)\sgpp(r_1,\oh)\sgppp(r_2,\oh')\rangle$. We
expressed these real-space correlators in terms of their Fourier
space analogue which take the form of
$C_l^{\sg\sgp\sgpp,\sgppp}(r_2,r_1)$ and
$C_l^{\sg\sgp,\sgpp\sgppp}(r_2,r_1)$.

The statistical descriptors we have presented here will provide
particularly useful tools for the study of non-Gaussianity in
alternative theories of gravity \citep{Ber04}; which is one of the
important science drivers for the future generations of weak lensing
surveys. We plan to present detailed results elsewhere in future.
The various analytical approximations, such as the Born
approximations linking the observed shear or convergence bispectrum
(or trispectrum) have been studied with some rigor
\citep{Ham02,ShapiCoo06}. The effect of source clustering, as well
as lens and source overlap - which also introduce corrections  -
have also been studied with some detail. We plan to extend these
studies for the power spectra presented here elsewhere.

The estimators  we have worked with can be generalized further to
take into account optimum weighting. This will involve inverse
covariance weighting the observed field harmonics. This will then be
useful in constructing a {\it matched filtering} version of the
estimator used here to fine tune the study of primordial as well as
gravity induced non-Gaussianity. While previous ray tracing
simulation of weak lensing used a flat-sky approach \citep{JSW00},
simulations are now available for the entire observed sky
\citep{Tey09}. This will provide an opportunity to test the
analytical results presented here. A detailed discussion will be
presented elsewhere (Munshi et al. 2010, in preparation).

\section{Acknowledgements}
\label{acknow} The initial phase of this work was completed when DM
was supported by a STFC rolling grant at the Royal Observatory,
Institute for Astronomy, Edinburgh. DM also acknowledges support
from STFC standard grant ST/G002231/1 at School of Physics and
Astronomy at Cardiff University where this work was completed. AC
and JS acknowledge support from NSF AST-0645427 and NASA NNX10AD42G.

\bibliography{paper.bbl}

\end{document}